\documentclass[pra,twocolumn, superscriptaddress]{revtex4-2}
\usepackage{graphicx}
\usepackage{epsfig}
\usepackage{amsfonts}
\usepackage{amsmath}
\usepackage{amssymb}
\usepackage{dsfont}
\usepackage{amsthm}% in order to use a black box at the end of proofs
\usepackage{color}
\usepackage{mathrsfs}
\usepackage{enumitem}
\usepackage[colorlinks=true,linkcolor=blue,citecolor=blue,urlcolor=blue]{hyperref}
\usepackage{comment}
\usepackage{braket}
\usepackage{physics}
\usepackage{hhline}
\usepackage{multirow}
\usepackage{soul}

\begin{document}

%%%%%%%%%%%%%%%%%%%%%%%%%%%%%%%%%%%%%%%%%%%%%%%%%%%%%%%%%%%%%%%%%%%

\title{Loophole-free Bell tests with randomly chosen subsets of measurement settings}

%%%%%%%%%%%%%%%%%%%%%%%%%%%%%%%%%%%%%%%%%%%%%%%%%%%%%%%%%%%%%%%%%%%

\author{Jaskaran Singh}
\email{jaskaran@us.es}
\affiliation{Departamento de F\'{\i}sica Aplicada II, Universidad de Sevilla, 41012 Sevilla, Spain}

\author{Ad\'an Cabello}
\email{adan@us.es}
\affiliation{Departamento de F\'{\i}sica Aplicada II, Universidad de Sevilla, 41012 Sevilla, Spain}
\affiliation{Instituto Carlos~I de F\'{\i}sica Te\'orica y Computacional, Universidad de Sevilla, 41012 Sevilla, Spain}

%%%%%%%%%%%%%%%%%%%%%%%%%%%%%%%%%%%%%%%%%%%%%%%%%%%%%%%%%%%%%%%%%%%

\begin{abstract}
There are bipartite quantum nonlocal correlations requiring very low detection efficiency to reach the loophole-free regime but that need too many measurement settings to be practical for actual experiments. This leads to the general problem of what can be concluded about loophole-free Bell nonlocality if only a random subset of these settings is tested. Here we develop a method to address this problem. We show that, in some cases, it is possible to detect loophole-free Bell nonlocality by testing only a small random fraction of the settings. The consequence is a higher detection efficiency. The method allows for the design of loophole-free Bell tests in which, given a quantum correlation that violates a Bell inequality, one can calculate the minimum fraction of contexts needed to reach the detection-loophole-free regime. The results also enforce a different way of thinking about how local realistic models or classical communication can be used to simulate quantum nonlocal correlations, as it shows that the amount of resources that are needed can be made arbitrarily large simply by considering more contexts.
\end{abstract}

%%%%%%%%%%%%%%%%%%%%%%%%%%%%%%%%%%%%%%%%%%%%%%%%%%%%%%%%%%%%%%%%%%%

\maketitle

%%%%%%%%%%%%%%%%%%%%%%%%%%%%%%%%%%%%%%%%%%%%%%%%%%%%%%%%%%%%%%%%%%%

\section{Introduction}

%%%%%%%%%%%%%%%%%%%%%%%%%%%%%%%%%%%%%%%%%%%%%%%%%%%%%%%%%%%%%%%%%%%

\subsection{Motivation} 

%%%%%%%%%%%%%%%%%%%%%%%%%%%%%%%%%%%%%%%%%%%%%%%%%%%%%%%%%%%%%%%%%%%

Quantum advantage, based on the violation of Bell inequalities~\cite{B64, NDN22,ZVR22,LZZ22, PGT23, LAL23, PAM10, SB20, AM16, SZB21}, requires ensuring that the observed correlations cannot be simulated with local realistic models~\cite{B64, B66,FC72,WJS98}. This demands a detection efficiency above a certain threshold $\eta_{\rm crit}$~\cite{P70} that depends on the nonlocal correlations. Recently~\cite{MCB22,XSS23}, several bipartite quantum correlations between high-dimensional quantum systems have been identified requiring $\eta_{\rm crit}$ smaller than those needed for qubits~\cite{GM87,E93}
and ququarts~\cite{VPB10}, which are the systems used in the detection-loophole-free Bell tests performed so far \cite{RKM01,GMR13,CMK13,SMC15,GVW15,HBD15,RBG17,HZL22}.
However, for all these cases, achieving a small detection efficiency requires a large number of settings. 
For example, to achieve $\eta_{\rm crit} = 0.510$ in Ref.~\cite{MCB22}, each party has to measure $2^{8}$ settings, to achieve $\eta_{\rm crit} = 0.324$ in Ref.~\cite{XSS23}, each party has to measure $2^{30}$ settings. These correlations need too many settings to be useful for actual experiments.

This begs the question of whether it is possible to detect loophole-free Bell nonlocality when the parties only randomly choose a fraction of these settings.

%%%%%%%%%%%%%%%%%%%%%%%%%%%%%%%%%%%%%%%%%%%%%%%%%%%%%%%%%%%%%%%%%%%

\subsection{Aim}

%%%%%%%%%%%%%%%%%%%%%%%%%%%%%%%%%%%%%%%%%%%%%%%%%%%%%%%%%%%%%%%%%%%

In a bipartite Bell experiment, a measurement context is one of the possible pairs of local settings used to evaluate the Bell inequality. 
Here, our aim is to investigate whether the parties can detect loophole-free Bell nonlocality with a certain confidence level using only a randomly chosen strict subset of the measurement contexts. If they can, we want to know how large $\eta$ must be to reach the loophole-free regime. Reciprocally, for a given detection efficiency, we want to know the minimum fraction of contexts needed to detect loophole-free Bell nonlocality. In particular, we want to elucidate whether this approach is useful when applied to the correlations in~\cite{MCB22,XSS23} and similar examples that may appear in the future.

%%%%%%%%%%%%%%%%%%%%%%%%%%%%%%%%%%%%%%%%%%%%%%%%%%%%%%%%%%%%%%%%%%%

\subsection{Structure}

%%%%%%%%%%%%%%%%%%%%%%%%%%%%%%%%%%%%%%%%%%%%%%%%%%%%%%%%%%%%%%%%%%%

The paper is organized as follows. 
In Sec.~\ref{sec:gen_strategy} we develop a general method to obtain the fraction of (randomly chosen) contexts needed to detect loophole-free Bell nonlocality with a given confidence value, provided we are given a correlation that violates a Bell inequality, and know the experimentally achievable detection efficiency $\eta_{\rm expt}$ (which has to be greater than the threshold detection efficiency $\eta_{\rm crit}$ in order to observe a violation). For that, we analyze the case of a correlation that violates a general bipartite Bell inequality and construct an estimator of the value of the Bell parameter as a function of the fraction of measurement contexts. Then, by using Chebyshev's inequality~\cite{T67}, we bound the minimum fraction of measurement contexts required.

In Secs.~\ref{sec:PNP_strategy} and~\ref{sec:graph_strategy} we apply the method to the quantum correlations maximally violating the so-called penalized $N$-product (PNP) Clauser-Horne-Shimony-Holt (CHSH) Bell inequalities~\cite{MCB22} and to some graph-theoretic Bell inequalities~\cite{XSS23}, respectively. We focus on these quantum correlations and Bell inequalities because they allow, with imperfect detection efficiency, us to produce loophole-free nonlocality even with a fraction of measurement contexts. For these cases, we also obtain how the detection efficiency depends on the fraction of measurement contexts. In Sec.~\ref{sec:conc} we summarize the pros and cons of the approach, explain why it offers a way to design loophole-free Bell tests, and discuss the implications of our results for the attempts to simulate quantum nonlocality with classical communication or local realistic models.

%%%%%%%%%%%%%%%%%%%%%%%%%%%%%%%%%%%%%%%%%%%%%%%%%%%%%%%%%%%%%%%%%%% 

\section{Loophole-free Bell violation using a subset of the measurement settings} 
\label{sec:gen_strategy}

%%%%%%%%%%%%%%%%%%%%%%%%%%%%%%%%%%%%%%%%%%%%%%%%%%%%%%%%%%%%%%%%%%% 

\subsection{Method}

%%%%%%%%%%%%%%%%%%%%%%%%%%%%%%%%%%%%%%%%%%%%%%%%%%%%%%%%%%%%%%%%%%% 

Consider two spatially separated parties, Alice and Bob, each of them performing measurements on a subsystem of a composite system. Let us denote Alice's measurement setting by $x$ and Bob's measurement setting by $y$. Let us denote the outcome of Alice's measurement by $a$ and the outcome of Bob's measurement by $b$. A Bell inequality is a bound on the linear combination of joint conditional probabilities $p(a,b|x,y)$ for the outcomes of different combinations of measurement settings of Alice and Bob. Specifically, a Bell inequality is an expression of the form
\begin{equation}
 \beta = \sum_{a, b, x, y}c_{a, b}^{x, y} p(a, b|x, y) = \sum_{j=1}^{M} \beta_j \leq C,
 \label{eq:bell}
\end{equation}
where $j = \left(x, y\right)$ is the measurement context (hereafter simply called context) corresponding to the settings $x$ and $y$, and thus $\beta_j = \sum_{a, b}c_{a, b}^{j}p(a, b|j)$, $M$ is the total number of contexts, and $C$ is the maximum achievable value by local realistic models. 
For example, in the CHSH Bell inequality \cite{CHS69}, $c_{a, b}^{x, y} = \delta_{a \oplus b, xy}$, $M = 4$, and $C = 3$.

Alice and Bob randomly select a subset of contexts of cardinality $L$ and evaluate the corresponding joint probability distributions. In principle, we assume that $L < M$. However, as we will see, not all nonlocal correlations allow us to certify nonlocality using only a subset of settings.
Now Alice and Bob want to estimate the value of $\beta$, that is, they want to obtain the most likely value of $\beta$ (whose detailed calculation requires measuring all contexts) from the probabilities of contexts that have been chosen.

For each of the chosen contexts, Alice and Bob evaluate the corresponding $\beta_j$. We assume that, for each of the contexts chosen, Alice and Bob conduct $K$ rounds of the experiment. This allows them to determine $\beta_j$ up to some finite precision $\epsilon'$ and probability of failure $\delta'$. Once all the terms $\beta_j$ are evaluated, Alice and Bob can estimate $\beta$ with some finite precision $\epsilon$ and probability of failure $\delta$. For the remainder of this paper, we take the precision $\epsilon = \beta - C$. Moreover, it should be noted that the value of $\delta$ can be chosen by setting the confidence level of the test~\cite{H18}. Here, the confidence level is a measure of how sure one is about the results of the test. It is generally stated in terms of the standard deviation $\sigma$ of a normal distribution, e.g., $4\sigma$, $5\sigma$, and $6\sigma$. As the name suggests, a higher confidence level indicates a lower probability of failure. While it is possible to select higher values, for the remainder of this paper we choose to have a $4\sigma$ level of confidence, corresponding to $\delta = 0.000\,03$, as we find that it is sufficient to showcase our results.

To estimate $\beta$, Alice and Bob can proceed as follows. They select $j\in \lbrace 1, \ldots, M\rbrace$ at random with probability $p(j) = \frac{1}{M}$. Here, for simplicity, we will assume a uniform probability distribution for $p(j)$. However, $p(j)$ can be tailored according to $\beta$ and may be not uniform. For simplicity, we will also assume that it is possible to evaluate $\beta_j$ with infinite precision, i.e., that $K \rightarrow \infty$. The case of finite precision will be discussed later. Then an estimator of $\beta$ is
\begin{equation}
 X = M \beta_j.
\end{equation}
By construction, $\langle X\rangle = \beta$, where the mean value is averaged over the randomly selected contexts $j$. It should be noted that this is not an average over the different experimental rounds of the Bell experiment.

Then the parties choose $L$ contexts. If the first context is context $p$, then we define $X_1=M \beta_p$; if the second context is context $q \neq p$, then $X_2=M \beta_q$; etc. Each $X_l$ is an estimator of the value of $\beta$. Let $Y$ be the average value of these estimators, that is,
\begin{equation}
\begin{aligned}
 Y &= \frac{1}{L} \sum_{l = 1}^{L} X_l\\
 &= \frac{M}{L} \sum_{i} \beta_{i},
\end{aligned}
\label{eq:average}
\end{equation}
where the second sum also has $L$ terms. The variance of each $X_l$ can also be bounded if we consider a uniform probability distribution $p(j) = \frac{1}{M}$. We have 
\begin{equation}
\begin{aligned}
    \text{Var}(X_l) &= \langle X^2_l \rangle - \langle X_l\rangle^2\\
    &=\sum_j p(j) M^2 \beta^2_j - \langle X_l \rangle^2\\
    &\leq M \sum_j \beta^2_j\\
    &\leq M \sum_j \beta_j\\
    &=M\beta,
\end{aligned}
\label{eq:variance}
\end{equation}
where the third equation is obtained using $\langle X_l \rangle^2 \geq 0$. We perform a further simplification to obtain the last inequality by noting that, for the Bell inequalities we consider here in the paper, $\beta_j \leq 1$. From Eqs.~\eqref{eq:average} and ~\eqref{eq:variance} we obtain Var$(Y) \leq \frac{M\beta}{L}$. Therefore, using Chebyshev's inequality~\cite{T67}, 
\begin{equation}
 p\left(|Y - \beta| \geq \lambda\sqrt{\frac{M\beta}{L}}\right) \leq \frac{1}{\lambda^2},
 \label{eq:chebyshev}
\end{equation}
where $\lambda > 0$ is a real number.

Since we want $Y$ with probability of failure $\delta$ and error $\epsilon$, we take 
\begin{equation}
 \lambda = \frac{1}{\sqrt{\delta}}
\end{equation}
and 
\begin{equation}
\label{ele}
 L = \frac{M\beta}{\epsilon^2 \delta}
\end{equation} 
to obtain
\begin{equation}
p\left(|Y - \beta| \geq \epsilon\right) \leq \delta.
\end{equation}
Then the fraction of contexts needed is
\begin{equation}
 \nu = \frac{L}{M} = \frac{\beta}{\epsilon^2 \delta}.
 \label{eq:frac_contexts}
\end{equation}

We now deal with the fact that $K$, the number of rounds used to evaluate each of the terms $\beta_j$, must be finite. In this case, an estimator of $\beta_j$ is the average value
\begin{equation}
 B_j = \frac{1}{K}\sum_{k = 1}^{K}\beta_{j}^{(k)},
\end{equation}
where $\beta_{j}^{(k)}$ is the value of $\beta_{j}$ obtained in round $k$. Let $\delta'$ be the probability of failure and $\epsilon'$ the error in evaluating $\beta_{j}$. Then, applying Hoeffding's inequality~\cite{H63}, we have
\begin{equation}
 p\left(B_j - \beta_{j} \geq \epsilon'\right) \leq \delta',
\end{equation}
where 
\begin{equation}
\begin{aligned}
 \delta' &= \exp \left[-\frac{2 \epsilon'^2}{\sum_{k = 1}^{K}\left(\frac{\beta_j}{K}\right)^2}\right]\\
 &= \exp \left(-\frac{2 \epsilon'^2 K}{\beta_{j}^2}\right).
\end{aligned}
\end{equation}

This yields
\begin{equation}
 K \geq \frac{-\ln (\delta') \beta_{j}^2}{2 \epsilon'^2},
 \label{eq:number_of_measurements}
\end{equation}
that is, the minimum number of rounds needed to estimate $\beta_j$.

So far, we have obtained the minimum fraction of contexts needed for a given $\beta$ and thus for a given detection efficiency. Another interesting problem is to obtain the minimum detection efficiency needed for a given fraction of contexts. For that, notice that $\epsilon = \beta - C$, where $\beta$ is the value of the Bell parameter obtained when the detection efficiency is $\eta$. Substituting this in Eq.~\eqref{eq:frac_contexts} and rearranging terms, we obtain
\begin{equation}
    \frac{(\beta - C)^2}{\beta} = \frac{1}{\nu \delta}.
    \label{eq:eta_given_frac}
\end{equation}
In general, $\beta$ is a function of $\eta$, the maximum quantum value $Q$, and the local value $C$ of the Bell inequality. The exact form of the function depends on how the no-click events are treated (see Ref.~\cite{XSS23} for more details). However, for a given Bell inequality and a model of detection efficiencies, Eq.~\eqref{eq:eta_given_frac} can be solved for $\eta$ in terms of $\nu$. 

%%%%%%%%%%%%%%%%%%%%%%%%%%%%%%%%%%%%%%%%%%%%%%%%%%%%%%%%%%%%%%%%%%% 

\subsection{The method does not always work}

%%%%%%%%%%%%%%%%%%%%%%%%%%%%%%%%%%%%%%%%%%%%%%%%%%%%%%%%%%%%%%%%%%% 

It is important to emphasize that the method is only useful for some quantum correlations that violate a Bell inequality with imperfect detection efficiency. In many cases, give a quantum correlation that violates a Bell inequality with imperfect detection efficiency, the method only says that all the contexts are needed. For example, consider the CHSH Bell inequality written so it has local value $C = 3$ and maximum quantum violation $2 + \sqrt{2} \approx 3.414$. 
The critical detection efficiency in this case is $\eta_{\text{crit}} = 2(\sqrt{2} - 1) \approx 0.828$~\cite{GM87}. Let us suppose that the value of the CHSH Bell parameter is $\beta = 3.272$ (thus $\epsilon = 0.272$) and has been achieved with $\eta = 0.880$ and $V = 1$, where $V$ is the visibility of the quantum state.
Let us assume that $\delta = 0.000\,03$, which corresponds to a test with a $4\sigma$ level of confidence, where $\sigma$ is the standard deviation of a normal distribution. By taking such a small value of $\delta$ we ensure that the probability to erroneously identify a Bell violation is also very small. Then $L \geq 5\,896\,771$. However, the total number of contexts in the CHSH Bell inequality is $M=4$. Therefore, $L > M$ indicates that it is not possible to consider a strict subset of contexts and obtain a loophole-free Bell violation. 

%%%%%%%%%%%%%%%%%%%%%%%%%%%%%%%%%%%%%%%%%%%%%%%%%%%%%%%%%%%%%%%%%%%%%%%%%

\section{The PNP Bell inequalities using a subset of settings}
\label{sec:PNP_strategy}

Here we apply the tools described in Sec.~\ref{sec:gen_strategy} to the case of the PNP Bell inequalities of Ref.~\cite{MCB22}. 

%%%%%%%%%%%%%%%%%%%%%%%%%%%%%%%%%%%%%%%%%%%%%%%%%%%%%%%%%%%%%%%%%%%

\subsection{The PNP Bell inequalities}

%%%%%%%%%%%%%%%%%%%%%%%%%%%%%%%%%%%%%%%%%%%%%%%%%%%%%%%%%%%%%%%%%%%

Given a Bell inequality with local realistic bound $C$, its PNP version is the product of $n$ copies of that Bell inequality with an extra penalization term chosen to guarantee that the local realistic bound of the PNP version is $C^n$. 
More specifically, the PNP Bell inequality can be written as
\begin{equation}
 \beta_{\text{PNP}} = \sum_{\textbf{a},\textbf{b},\textbf{x},\textbf{y}}p(\textbf{a}, \textbf{b}|\textbf{x},\textbf{y})\prod^{n}_{i = 1} c_{a_i,b_i}^{x_i,y_i} - \kappa (A + B) \leq C^n,
 \label{eq:bell_PNP}
\end{equation}
where $\textbf{x} = (x_1,\ldots, x_n)$ is Alice's measurement setting (which can be seen as $n$ measurement settings, one for each copy of the original Bell inequality), $\textbf{y} = (y_1,\ldots, y_n)$ is Bob's measurement setting, $\textbf{a} = (a_1,\ldots, a_n)$ is Alice's outcome (which can be seen as $n$ outcomes, one for each copy of the original Bell inequality) with $a_i \in \left[m\right] ~\forall i\in\lbrace1, 2,\ldots, n\rbrace$, and $\textbf{b} = (b_1,\ldots, b_n)$ is Bob's outcome with $b_i \in \left[m\right] ~\forall i\in\lbrace1, 2,\ldots, n\rbrace$. In addition,
$\kappa = 2^{n-1} (\Sigma_n - C^n)$, where 
$C$ is the maximum local bound of the original Bell inequality and $\Sigma_n$ is the algebraic bound of Eq.~\eqref{eq:bell_PNP} without the penalization term . In addition,
\begin{subequations}
\begin{align}
 A &= \sum_{i = 1}^{n}\sum_{\textbf{x}}\sum_{\mathbf{x}\neq \mathbf{x’} | x'_i=x_i }\sum_{a_i = 0}^{m-2}|p(a_i|\textbf{x}) - p(a_i|\textbf{x}')|,\\
 B &= \sum_{i = 1}^{n}\sum_{\textbf{y}}\sum_{\textbf{y}\neq \textbf{y}' | y'_i = y_i} \sum_{b_i = 0}^{m-2}|p(b_i|\textbf{y}) - p(b_i|\textbf{y}')|,
\end{align}
\end{subequations}
where the third summation is taken over $\textbf{x}$ and $\textbf{x}'$ such that they only match on the $i$th element, but are different on all other elements, and similarly for the sum over $\textbf{y}'$. Let $Q$ be the maximum quantum value of the original Bell inequality. Then the maximum quantum value of the corresponding PNP Bell inequality is simply $Q^n$.

%%%%%%%%%%%%%%%%%%%%%%%%%%%%%%%%%%%%%%%%%%%%%%%%%%%%%%%%%%%%%%%%%%%%%%%%%
% Fig. 1
%%%%%%%%%%%%%%%%%%%%%%%%%%%%%%%%%%%%%%%%%%%%%%%%%%%%%%%%%%%%%%%%%%%%%%%%%

\begin{figure}
 \centering
 \includegraphics[scale = 0.6]{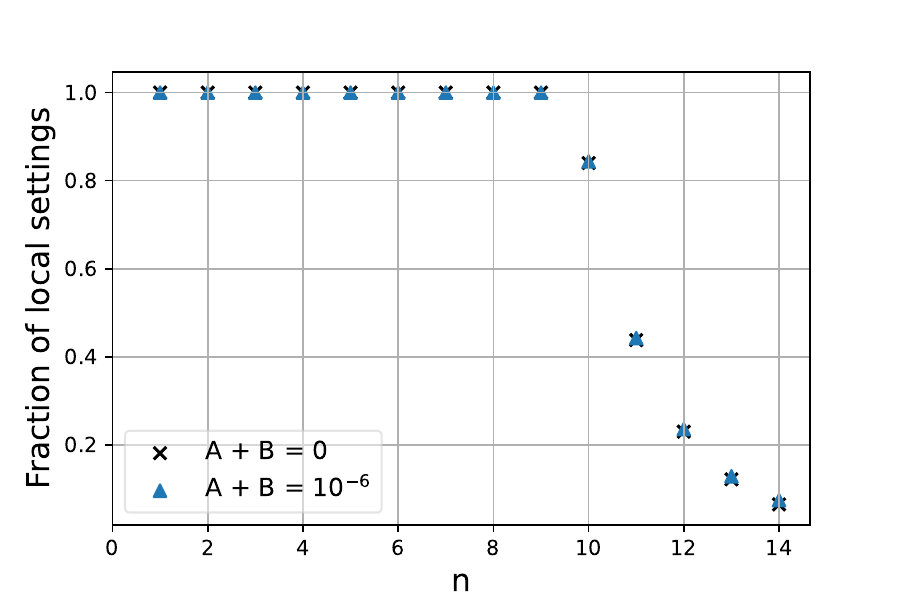}
 \caption{Fraction of the settings required for a loophole-free test of the PNP CHSH Bell inequalities, as a function of $n$, the number of CHSH Bell inequalities conducted in parallel.}
 \label{fig:pnp_bell}
\end{figure}

%%%%%%%%%%%%%%%%%%%%%%%%%%%%%%%%%%%%%%%%%%%%%%%%%%%%%%%%%%%%%%%%%%%

\subsection{The PNP CHSH Bell inequalities with a subset of settings}

%%%%%%%%%%%%%%%%%%%%%%%%%%%%%%%%%%%%%%%%%%%%%%%%%%%%%%%%%%%%%%%%%%%
 
We focus on the PNP Bell inequalities 
constructed by taking $n$ copies of the CHSH Bell inequality. In this case, $c_{a_i, b_i}^{x_i, y_i} = \delta_{a_i \oplus b_i, x_iy_i}$ $\forall a_i, b_i, x_i, y_i$, $C = 3$, and $Q = 2 + \sqrt{2}$. There are $2^n$ measurement settings per party and the total number of contexts is $M = 4^{n}$. For the time being, we will assume that $A + B = 0$. Later on, we will consider the case of $A + B \neq 0$. This scenario can also be visualized as having $n$ bipartite two-qubit states, on each of which we perform a copy of the CHSH Bell test.

The parties randomly select $L$ contexts. Each of them is tested $K$ times. Applying the tools in Sec.~\ref{sec:gen_strategy}, we obtain 
the fraction of local settings that each party must choose randomly to obtain a loophole-free Bell violation of the PNP CHSH Bell inequality such that the probability of failure is $\delta = 0.000\,03$ and under the assumption that the visibility of each bipartite two-qubit state is $V = 0.9$. The results are shown in Fig.~\ref{fig:pnp_bell}. There we observe that the fraction is smaller than $1$ only after $n = 9$.

Next we calculate the minimum detection efficiency $\eta_{\nu}$ for a loophole-free Bell violation as a function of the fraction of contexts chosen, $\nu = \frac{L}{M}$. In order to do so, we model detection inefficiencies as follows. We bin the no clicks to one of the outcomes (always the same one). Then
\begin{equation}
 \beta_{\text{PNP}} = \eta^2 Q^n + \eta (1 - \eta)(P_A^n + P_B^n) + (1 - \eta)^2 C^n,
 \label{eq:detect_pnp}
\end{equation}
with
\begin{subequations}
\begin{align}
 Q &= \sum_{a, b, x, y} c_{a, b}^{x, y} \text{tr}\left(\mathcal{A}_a^x \otimes \mathcal{B}_b^y \rho_{AB}\right),\\
 P_A &= \sum_{a, b, x, y} c_{a, b}^{x, y} \text{tr} \left(\mathcal{A}_a^x \rho_A\right),\\
 P_B &= \sum_{a, b, x, y} c_{a, b}^{x, y} \text{tr} \left(\mathcal{B}_b^y \rho_B\right),
\end{align}
\end{subequations}
where the state shared by Alice and Bob is of the form $\bigotimes_{i = 1}^{n}\rho_{AB}$, and $\bigotimes_{i = 1}^{n}\mathcal{A}_{a_i}^{x_i}$ and $\bigotimes_{i = 1}^{n}\mathcal{B}_{b_i}^{y_i}$ are the elements of the positive-operator-valued-measures. We assume that $\mathcal{A}_{a_i}^{x_i} = \mathcal{A}_{a}^{x}$ and $\mathcal{B}_{b_i}^{y_i} = \mathcal{B}_{b}^{y} \forall i$.

\begin{figure}
 \centering
 \includegraphics[scale = 0.56]{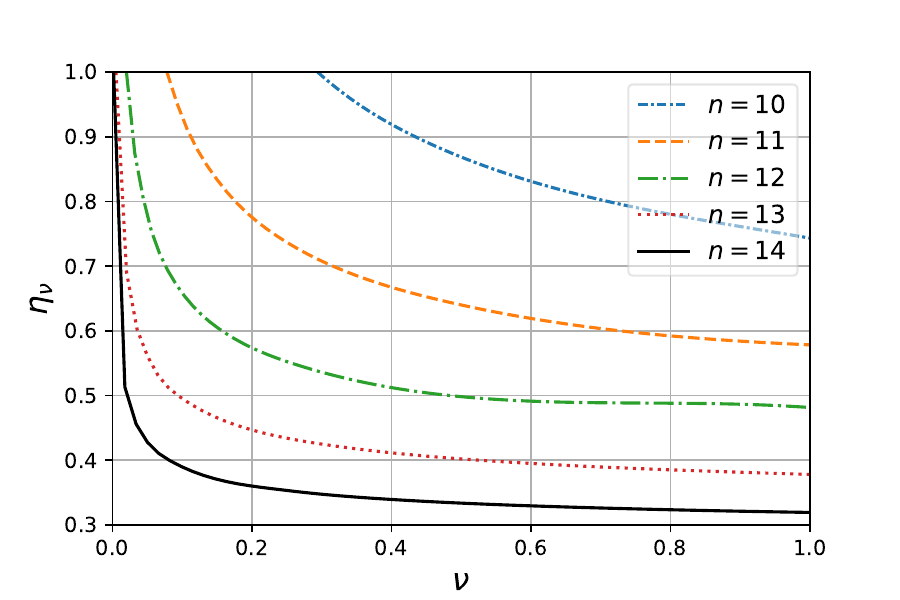}
 \caption{Minimum detection efficiency $\eta_{\nu}$ needed to reach the loophole-free regime as a function of the fraction $\nu$ of the contexts for the PNP CHSH Bell inequality with different values of $n$. The threshold detection efficiencies for $n = 10, 11, 12, 13$, and $14$ are $\eta_{\rm crit} = 0.43, 0.38, 0.34, 0.31$, and $0.28$, respectively.}
 \label{fig:pnp_frac_eta}
\end{figure}

Figure~\ref{fig:pnp_frac_eta} shows the minimum detection efficiency $\eta_{\nu}$ needed for a loophole-free Bell violation, when using a fraction $\nu$ of all the contexts in the PNP CHSH Bell inequality for $n = 10, 11, 12, 13$, and $14$, for $\delta = 0.000\,03$ and $V = 1$. Notice that, in all cases, $\nu=1$ occurs before the respective $\eta_{\rm crit}$ is reached ($\eta_{\rm crit}$ for $n = 10, 11, 12, 13$, and $14$ are $0.43, 0.38, 0.34, 0.31$, and $0.28$, respectively). This implies that, for very low Bell violations (with detection efficiency close to the critical value) and a given confidence level (determined by $\delta$), our method cannot guarantee a loophole-free Bell violation unless all contexts are measured. By choosing higher values of $\delta$ (low confidence) it is possible to reach detection efficiencies close to the critical values. 

%%%%%%%%%%%%%%%%%%%%%%%%%%%%%%%%%%%%%%%%%%%%%%%%%%%%%%%%%%%%%%%
% Fig. 2
%%%%%%%%%%%%%%%%%%%%%%%%%%%%%%%%%%%%%%%%%%%%%%%%%%%%%%%%%%%%%%%

%%%%%%%%%%%%%%%%%%%%%%%%%%%%%%%%%%%%%%%%%%%%%%%%%%%%%%%%%%%%%%%

Finally, we analyze the case when $A + B \neq 0$. The value of this sum cannot be very large; otherwise a violation will not be observed. This is because $\kappa$ increases exponentially with $n$. As an example, consider $A + B = 10^{-6}$, for which the parties can observe a violation up to $n = 14$. We use this particular value of $A + B$ so that some violation of the PNP Bell inequality can be observed for at most $n = 14$ (which corresponds to the last data point in Fig.~\ref{fig:pnp_bell}). It can be seen that, by construction of Eq.~\eqref{eq:bell_PNP}, for larger values of $A + B$, no violation can be observed for $n = 14$ due to an exponentially large value of $\kappa$. 

Then, the maximum quantum violation for the PNP Bell inequality is $\beta^n - 10^{-6}\kappa$. As it can be seen from Fig.~\ref{fig:pnp_bell}, the number of random local settings chosen by either of the parties does not increase significantly from the case when $A + B = 0$. Therefore, it is possible for Alice and Bob to detect nonlocality even when $A + B \neq 0$.

%%%%%%%%%%%%%%%%%%%%%%%%%%%%%%%%%%%%%%%%%%%%%%%%%%%%%%%%%%%%%%%%%%%%%%%

\section{Graph-theoretic Bell inequalities using a subset of settings}
\label{sec:graph_strategy}

%%%%%%%%%%%%%%%%%%%%%%%%%%%%%%%%%%%%%%%%%%%%%%%%%%%%%%%%%%%%%%%%%%%

Here we apply the tools described in Sec.~\ref{sec:gen_strategy} to the case of the graph-theoretic Bell inequalities of Ref.~\cite{XSS23}.

%%%%%%%%%%%%%%%%%%%%%%%%%%%%%%%%%%%%%%%%%%%%%%%%%%%%%%%%%%%%%%%%%%%

\subsection{The graph-theoretic Bell inequalities}

%%%%%%%%%%%%%%%%%%%%%%%%%%%%%%%%%%%%%%%%%%%%%%%%%%%%%%%%%%%%%%%%%%%

Consider two separated parties, Alice and Bob, each of them having access to a set of measurement settings corresponding in quantum mechanics to projectors $\lbrace \Pi_i\rbrace$, each of them having two outcomes, and such that the graph $G$ describes the relations of orthogonality between the members of $\lbrace \Pi_i\rbrace$: Each element of $\lbrace \Pi_i\rbrace$ is represented by a vertex of $G$ and orthogonal projectors correspond to adjacent vertices. Then the corresponding graph-theoretic Bell inequality can be written as
\begin{equation}
\begin{aligned}
 \beta_G &= \sum_{i \in \mathcal{V}} p(\Pi_i^A = \Pi_i^B = 1) \\
 &\quad - \sum_{(i, j) \in \mathcal{E}}\frac{1}{2\Xi}\left[p(\Pi_i^A = \Pi_j^B = 1) + p(\Pi_j^A = \Pi_i^B = 1)\right]\\
 &\leq C,
\end{aligned}
 \label{eq:graph_bell}
\end{equation}
where $\mathcal{V}$ is the vertex set of $G$, $\mathcal{E}$ is the edge set of $G$, $\Xi$ is the Xi number of $G$ (which, for simplicity, can be taken to be $1$), and 
$p(\Pi_i^A = \Pi_j^B = 1)$ is the probability for Alice and Bob to obtain the outcome $1$ when they measure $\Pi_i^A$ and $\Pi_j^B$, respectively.
Interestingly, both the maximum local realistic value $C$ and the maximum quantum value $Q$ can be related to properties of $G$. Specifically, $C$ is the independence number of $G$ and $Q$ is $\frac{|\mathcal{V}|}{\xi}$, where $\xi$ is the orthogonal rank of the graph $G$ (see Ref.~\cite{XSS23} for details). 

%%%%%%%%%%%%%%%%%%%%%%%%%%%%%%%%%%%%%%%%%%%%%%%%%%%%%%%%%%%%%%%%%%%

\subsection{The graph-theoretic Bell inequalities with a subset of settings}

%%%%%%%%%%%%%%%%%%%%%%%%%%%%%%%%%%%%%%%%%%%%%%%%%%%%%%%%%%%%%%%%%%%

For any graph $G$, the number of settings per party in $\beta_G$ in \eqref{eq:graph_bell} is $|\mathcal{V}|$ and the total number of contexts is $M = |\mathcal{V}| + 2|\mathcal{E}|$. 
To estimate $\beta_G$, Alice and Bob choose settings $i$ and $j$ from the set $\mathcal{V}$ with the nonuniform probability distribution
\begin{equation}
 p(i, j) = \begin{cases}
 \frac{|\mathcal{V}|}{|\mathcal{V}| + 2 |\mathcal{E}|} \quad \text{for } i = j\\
 \frac{2|\mathcal{E}|}{|\mathcal{V}| + 2|\cal{E}|} \quad \text{for }(i, j)\in \cal{E}\\
 0 \quad\quad\quad\quad \text{otherwise}.
 \end{cases}
\end{equation}
Then the estimator is
\begin{equation}
 X_{i, j} = \begin{cases}
 \frac{1}{p(i, j)}\beta_{i,j} \quad \text{for } i = j\\
 \frac{-1}{p(i, j)}\beta_{i,j} \quad \text{for } (i, j) \in \mathcal{E}\\
 0 \quad\quad\quad\quad \text{otherwise},
 \end{cases}
\end{equation}
where $\beta_{i, j} = p(\Pi_i^A = \Pi_j^B = 1)$. It can be seen that the mean value of the estimator $X_{i, j}$, when averaged over the variables $i$ and $j$, is simply $\langle X_{i, j} \rangle = \beta_G$. When $(i, j) \in \mathcal{E}$ or $i = j$, the average is simply over the randomly selected contexts. Again, it should be noted here that this is not an average over the different rounds of the Bell experiment.

Then Alice and Bob choose $L$ contexts according to the distribution $p(i, j)$ to evaluate the corresponding $L$ estimators $X_{i,j}$. In this case, we apply Hoeffding's inequality instead of Chebyshev's inequality to evaluate the number of contexts required. The reason is that Chebyshev's inequality requires evaluating the variance of each estimator in terms of the Bell value, which is not possible in this case because of the inherent asymmetry in the Bell inequality. Instead, we can use the fact that each estimator $X_{i, j}$ is a function of a probability distribution and can thus be bounded. 

Following Eq.~\eqref{eq:average}, let $Y$ be the average value of the estimators corresponding to the $L$ contexts chosen by the parties.
Since $Y$ is a sum of independent random variables $\frac{X_l}{L}$, we can apply Hoeffdings's inequality to bound its value. In order to do so, we can bound each of the estimators by noting that their achievable maximum and minimum values are $\frac{(|\mathcal{V}| + 2|\mathcal{E}|)}{|\mathcal{V}|}$ and $\frac{-(|\mathcal{V}| + 2|\mathcal{E}|)}{2|\mathcal{E}|}$, respectively. Using Hoeffding's inequality, we have 
\begin{equation}
 p(Y - \beta_G \geq \epsilon) \leq \delta.
\end{equation}

For a fixed precision $\epsilon$ and probability of failure $\delta$, the number of contexts that Alice and Bob would need to evaluate is
\begin{equation}
 L = \frac{-\ln(\delta) \left(|\mathcal{V}| + 2|\mathcal{E}|\right)^4}{8\epsilon^2 |\mathcal{E}|^2|\mathcal{V}|^2}.
\end{equation}

Now we calculate the minimum detection efficiency $\eta_\nu$ needed for loophole-free Bell violation as a function of the fraction $\nu$ of the contexts. In order to model the detection inefficiency, we bin the no clicks to the outcome $0$. This outcome is chosen because no terms corresponding to it appear in the Bell inequality~\eqref{eq:graph_bell}. This simplifies the evaluation of the critical detection efficiency, which, in this case, is now only a function of the maximum local realistic and quantum values.
Then the minimum detection efficiency is
\begin{equation}
 \eta_\nu = \left[\frac{1}{Q}\left(\frac{-\ln \delta \left(|\mathcal{V}| + 2|\mathcal{E}|\right)^3}{2 \nu |\mathcal{V}|^2 |\mathcal{E}|^2}\right)^{1/2} + \frac{C}{Q}\right]^{1/2}.
\end{equation}

Next we consider nine different correlations associated with the maximum quantum violation of a different Bell inequality, each of them corresponding to a different graph. The details of these graphs can be found in Ref.~\cite{XSS23} but are not relevant for our purposes. For each case, we calculate the fraction of contexts needed for a given detection efficiency $\eta$. The results are presented in Table~\ref{tab:graph_theoretic}. 
As it can be seen in Table~\ref{tab:graph_theoretic}, in all the cases except the last correlation, by randomly selecting a small strict subset of the contexts, the parties can claim loophole-free Bell nonlocality with a very small probability of failure. In the last correlation presented in Table~\ref{tab:graph_theoretic}, the parties will have to select all contexts for all $\eta \geq \eta_{\rm crit} = 0.912$. 

%%%%%%%%%%%%%%%%%%%%%%%%%%%%%%%%%%%%%%%%%%%%%%%%%%%%%%%%%%%%%%%%%%%
% Table II
%%%%%%%%%%%%%%%%%%%%%%%%%%%%%%%%%%%%%%%%%%%%%%%%%%%%%%%%%%%%%%%%%%%
\setlength{\tabcolsep}{7pt}
\begin{table}
 \centering
 \caption{Fraction $\nu$ of contexts needed to detect loophole-free Bell nonlocality for different values of the detection efficiency $\eta$, for the correlations violating some of the graph-theoretic Bell inequalities. Bell inequality indicates the Bell inequality considered, $d$ is the dimension of each of the local quantum systems, and $M$ is the total number of contexts. Here we assume the visibility of the quantum states $V=1$ and $\delta = 0.000\,03$.} 
 
 \begin{tabular*}{\linewidth}{l c c c c}
 \hline
 \hline
Bell inequality & $d$ & $M$ & $\eta$ & $\nu$ \\
\hline
 
%%%%%%%%%%%%%%%%%%%%%%%%%%%%%%%%%%%%%%%%%%%%%%%%%%%%%%%%%%%%%%%%%%%%%%%

$Y_{44}$ & $44$ & $4.62 \times 10^{24}$ & $0.163$ & $1$\\
 
$Y_{44}$ & $44$ & $4.62 \times 10^{24}$ & $0.200$ & $7.01 \times 10^{-19}$\\
 
$Y_{44}$ & $44$ & $4.62 \times 10^{24}$ & $0.400$ & $7.01 \times 10^{-21}$\\

$Y_{44}$ & $44$ & $4.62 \times 10^{24}$ & $0.600$ & $1.12 \times 10^{-21}$\\

$Y_{44}$ & $44$ & $4.62 \times 10^{24}$ & $0.800$ & $3.31 \times 10^{-22}$\\

$Y_{44}$ & $44$ & $4.62 \times 10^{24}$ & $0.950$ & $1.62 \times 10^{-22}$\\

%%%%%%%%%%%%%%%%%%%%%%%%%%%%%%%%%%%%%%%%%%%%%%%%%%%%%%%%%%%%%%%%%%%%%%%

$Y_{36}$ & $36$ & $7.79 \times 10^{19}$ & $0.260$ & $1$\\
 
$Y_{36}$ & $36$ & $7.79 \times 10^{19}$ & $0.400$ & $7.07 \times 10^{-16}$\\
 
$Y_{36}$ & $36$ & $7.79 \times 10^{19}$ & $0.600$ & $7.06 \times 10^{-17}$\\

$Y_{36}$ & $36$ & $7.79 \times 10^{19}$ & $0.800$ & $1.84 \times 10^{-17}$\\

$Y_{36}$ & $36$ & $7.79 \times 10^{19}$ & $0.950$ & $8.66 \times 10^{-18}$\\

%%%%%%%%%%%%%%%%%%%%%%%%%%%%%%%%%%%%%%%%%%%%%%%%%%%%%%%%%%%%%%%%%%%%%%%

$Y_{32}$ & $32$ & $3.22 \times 10^{17}$ & $0.326$ & $1$\\
 
$Y_{32}$ & $32$ & $3.22 \times 10^{17}$ & $0.400$ & $4.51 \times 10^{-13}$\\
 
$Y_{32}$ & $32$ & $3.22 \times 10^{17}$ & $0.600$ & $2.03 \times 10^{-14}$\\

$Y_{32}$ & $32$ & $3.22 \times 10^{17}$ & $0.800$ & $4.54 \times 10^{-15}$\\

$Y_{32}$ & $32$ & $3.22 \times 10^{17}$ & $0.950$ & $2.02 \times 10^{-15}$\\

%%%%%%%%%%%%%%%%%%%%%%%%%%%%%%%%%%%%%%%%%%%%%%%%%%%%%%%%%%%%%%%%%%%%%%%

$Y_{28}$ & $28$ & $1.34 \times 10^{15}$ & $0.407$ & $1$\\
 
$Y_{28}$ & $28$ & $1.34 \times 10^{15}$ & $0.600$ & $7.19 \times 10^{-12}$\\
 
$Y_{28}$ & $28$ & $1.34 \times 10^{15}$ & $0.800$ & $1.20 \times 10^{-12}$\\

$Y_{28}$ & $28$ & $1.34 \times 10^{15}$ & $0.950$ & $4.99 \times 10^{-13}$\\

%%%%%%%%%%%%%%%%%%%%%%%%%%%%%%%%%%%%%%%%%%%%%%%%%%%%%%%%%%%%%%%%%%%%%%%

$\mathcal{P}_4(\mathds{R})$ & $16$ & $8752320$ & $0.516$ & $1$\\
 
$\mathcal{P}_4(\mathds{R})$ & $16$ & $8752320$ & $0.600$ & $3.84 \times 10^{-4}$\\
 
$\mathcal{P}_4(\mathds{R})$ & $16$ & $8752320$ & $0.800$ & $2.40 \times 10^{-4}$\\

$\mathcal{P}_4(\mathds{R})$ & $16$ & $8752320$ & $0.950$ & $8.29 \times 10^{-5}$\\

%%%%%%%%%%%%%%%%%%%%%%%%%%%%%%%%%%%%%%%%%%%%%%%%%%%%%%%%%%%%%%%%%%%%%%%

$\mathcal{P}_3(\mathds{C})$ & $8$ & $341280$ & $0.730$ & $1$\\
 
$\mathcal{P}_3(\mathds{C})$ & $8$ & $341280$ & $0.750$ & $0.098$\\
 
$\mathcal{P}_3(\mathds{C})$ & $8$& $341280$ & $0.850$ & $0.002$\\

$\mathcal{P}_3(\mathds{C})$ & $8$ & $341280$ & $0.950$ & $6.17 \times 10^{-4}$\\

%%%%%%%%%%%%%%%%%%%%%%%%%%%%%%%%%%%%%%%%%%%%%%%%%%%%%%%%%%%%%%%%%%%%%%%

$\mathcal{P}_3(\mathds{R})$ & $8$ & $25440$ & $0.730$ & $1$\\
 
$\mathcal{P}_3(\mathds{R})$ & $8$ & $25440$ & $0.850$ & $0.072$\\

$\mathcal{P}_3(\mathds{R})$ & $8$ & $25440$ & $0.950$ & $0.019$\\

%%%%%%%%%%%%%%%%%%%%%%%%%%%%%%%%%%%%%%%%%%%%%%%%%%%%%%%%%%%%%%%%%%%%%%%

$\mathcal{P}_2(\mathds{C})$ & $4$ & $960$ & $0.894$ & $1$\\

$\mathcal{P}_2(\mathds{C})$ & $4$ & $960$ & $0.950$ & $0.668$\\

%%%%%%%%%%%%%%%%%%%%%%%%%%%%%%%%%%%%%%%%%%%%%%%%%%%%%%%%%%%%%%%%%%%%%%%

$\mathcal{P}_2(\mathds{R})$ & $4$ & $240$ & $0.912$ & $1$\\
 
%%%%%%%%%%%%%%%%%%%%%%%%%%%%%%%%%%%%%%%%%%%%%%%%%%%%%%%%%%%%%%%%%%%%%%%

\hline
\hline
\end{tabular*}
 \label{tab:graph_theoretic}
\end{table}

%%%%%%%%%%%%%%%%%%%%%%%%%%%%%%%%%%%%%%%%%%%%%%%%%%%%%%%%%%%%%%%%%%%

\section{Discussion}
\label{sec:conc}

%%%%%%%%%%%%%%%%%%%%%%%%%%%%%%%%%%%%%%%%%%%%%%%%%%%%%%%%%%%%%%%%%%%

\subsection{Pros and cons of the approach}

%%%%%%%%%%%%%%%%%%%%%%%%%%%%%%%%%%%%%%%%%%%%%%%%%%%%%%%%%%%%%%%%%%%

So far, we have shown that, for some bipartite correlations that violate a Bell inequality with imperfect detection efficiency and under the assumption that the
detection efficiency available is above the critical value, a small fraction of the contexts is enough to detect loophole-free Bell nonlocality with a given confidence. For example, in the PNP CHSH Bell inequality with $n = 13$, we have shown that the parties need only a fraction of $0.528$ of the contexts to detect loophole-free nonlocality. However, then the detection efficiency must be higher than the critical detection efficiency had the parties measured all contexts (specifically, it must be $\eta \geq 0.4$, while $\eta_{\rm crit} = 0.313$ when all contexts are measured). Moreover, the number of settings ($0.528 \times 10^{13}$) needed to measure that fraction of contexts is still too large to be practical.

Therefore, measuring only a fraction of the contexts is sometimes enough to detect loophole-free Bell nonlocality. However, this comes at the cost of a higher detection efficiency and, at least in the example, the reduction in the number of settings is not enough to be practical for standard experiments.

Then the question is what this approach useful for. In the following, we will argue that (i) it provides a way to design loophole-free Bell tests; (ii) it enforces a different way to look at the classical simulation of quantum correlations, as it shows that the amount of classical resources may depend on how large the fraction of random contexts is, which is a choice that can be modified during the experiment; and (iii) it may stimulate the search for correlations with many settings and low detection efficiency, to which no attention has been paid so far.

%%%%%%%%%%%%%%%%%%%%%%%%%%%%%%%%%%%%%%%%%%%%%%%%%%%%%%%%%%%%%%%%%%%

\subsection{Designing loophole-free Bell tests}

%%%%%%%%%%%%%%%%%%%%%%%%%%%%%%%%%%%%%%%%%%%%%%%%%%%%%%%%%%%%%%%%%%%

The method introduced here offers a different approach to the design of loophole-free Bell tests. Suppose that one can prepare the correlations needed to maximally violate the Bell inequality $Y_{32}$ in Table~\ref{tab:graph_theoretic} (for simplicity, we assume that $V=1$) and have in the laboratory a detection efficiency $\eta_{\rm expt} > \eta_{\rm crit}$, where $\eta_{\rm crit} = 0.326$. So far, the only option to detect loophole-free Bell nonlocality was evaluating all $M=3.22 \times 10^{17}$ contexts. The method offers an alternative. Suppose that $\eta_{\rm expt} \geq 0.400$. Then the method shows that a randomly chosen fraction of contexts $\nu \geq 4.51 \times 10^{-13}$ is enough to conclude loophole-free Bell nonlocality.
Similarly, if $\eta_{\rm expt} \geq 0.600$, then the fraction of contexts further reduces to $\nu \geq 2.03 \times 10^{-14}$, which can be further reduced to $\nu \geq 4.54 \times10^{-15}$ if $\eta_{\rm expt} \geq 0.800$, etc. 
This way, by knowing the detection efficiency, the method gives the minimum fraction $\nu$ of contexts needed. 

Even if $\nu$ is still too large to be practical, it is important to observe that the fact that the contexts are randomly chosen allows us to estimate the value of a Bell parameter that cannot be fully evaluated. For example, in all the examples considered, all the local measurements are equally difficult from an experimental point of view. For example, all the local measurements needed for the maximum quantum violation of the graph-theoretic Bell inequalities associated with the graphs $Y_d$ are represented by projectors that only differ in phases (see~\cite{XSS23} for details). Therefore, it is reasonable to expect that all contexts are equally affected by the experimental imperfections. Therefore, by measuring a sufficiently large random subset of contexts, a reliable estimate can be obtained of what would be obtained if a larger fraction were measured. This way, the correlations considered in this work can be used to estimate whether or not an experiment would reach the loophole-free regime if left running longer.

%%%%%%%%%%%%%%%%%%%%%%%%%%%%%%%%%%%%%%%%%%%%%%%%%%%%%%%%%%%%%%%%%%%

\subsection{Making classical simulations asymptotically impossible}

%%%%%%%%%%%%%%%%%%%%%%%%%%%%%%%%%%%%%%%%%%%%%%%%%%%%%%%%%%%%%%%%%%%

The tools introduced here also enforce a way to look at the cost of simulating Bell nonlocality {using local realistic models or classical communication~\cite{TB03,ZG19, BHQ15}. It should be noted that this cost of simulating nonlocal correlations is different from a computational cost, where, the cost may incorporate the memory and/or the time required to simulate the correlations. As an example, it is possible to computationally simulate the CHSH Bell nonlocal correlations without any high cost. However, this simulation is not possible using only a local realistic model. In a local realistic model that tries to simulate certain nonlocal correlations, the detection efficiency is also a target of the simulation (see, e.g., \cite{Larsson:1999PLA,Aerts:1999PRL, Cabello2009PRL}), that is, the local realistic model not only should reproduce the quantum statistics but should also simulate a fixed detection efficiency $\eta < \eta_{\rm crit}$. However, there are correlations that can be classically simulated for a fixed $\eta$ that cannot be simulated when more contexts are added. The problem is identifying them.

For example, consider a local realistic model which simulates $\eta = 0.9$ and the correlations maximally violating the Bell inequality in $\mathcal{P}_2(\mathds{R})$ for the $240$ contexts in Table~\ref{tab:graph_theoretic}. Such a local realistic model is possible because the simulated $\eta$ is smaller than the corresponding critical detection efficiency $\eta_{\rm crit} = 0.912$; otherwise the simulation would be impossible. Now note that the correlations maximally violating the $\mathcal{P}_2(\mathds{R})$ Bell inequality are a subset of the correlations maximally violating $\mathcal{P}_2(\mathds{C})$ (see \cite{XSS23} for details). If the parties decide to test more contexts until they cover all the $960$ contexts needed for the correlations maximally violating the $\mathcal{P}_2(\mathds{C})$ Bell inequality, then no local realistic model simulating $\eta = 0.9$ can also simulate the correlations. The reason is that simultaneously simulating the correlations for $\mathcal{P}_2(\mathds{C})$ and $\eta = 0.9$ is impossible since, in this case, $\eta_{\rm crit} = 0.894$ (see Table~\ref{tab:graph_theoretic}). 

However, it is not true that this is always the case. For example, the correlations maximally violating the $\mathcal{P}_3(\mathds{R})$ Bell inequality are a subset of the correlations maximally violating $\mathcal{P}_3(\mathds{C})$. However, the critical detection efficiency is the same in both cases (see Table~\ref{tab:graph_theoretic}).

Moreover, both examples above refer to nonlocal correlations in which we can identify a subset of them which is also nonlocal and we can compute the corresponding $\eta_{\rm crit}$. The problem is that identifying such subsets of correlations can be difficult. Nevertheless, in this work we have found correlations (all those in Fig.~\ref{fig:pnp_bell} and Table~\ref{tab:graph_theoretic}) for which a smaller randomly chosen fraction always requires a larger $\eta_{\rm crit}$ than a larger randomly chosen fraction. As mentioned, this is not true in general, even in the case when the parties are allowed to choose specific subsets. However, the power of our approach is that now the parties can calculate $\eta_{\rm crit}$ for any fraction of contexts. In all the cases in Fig.~\ref{fig:pnp_bell} and Table~\ref{tab:graph_theoretic} (except for the last correlation in Table \ref{tab:graph_theoretic}; this is why there is only a single row there), by simply increasing the fraction of contexts, the parties can make it impossible for a local realistic model to simulate the correlations.

This brings us to a final thought: Can the classical simulation succeed in reproducing the nonlocal correlations when the parties do not fix a specific number of contexts to measure prior to performing the Bell experiment? As an example, the parties can choose to terminate the experiment anytime after they have reached a sufficient number of contexts. In such a case, for not failing, the simulation should work for all contexts. There are probably cases in which the resources needed for the simulation rapidly tend to infinity as the number of all contexts in the Bell inequality increases. This question and this conjecture indicate that it would be interesting to investigate how fast these resources can grow with the number of contexts.

%%%%%%%%%%%%%%%%%%%%%%%%%%%%%%%%%%%%%%%%%%%%%%%%%%%%%%%%%%%%%%%%%%%

\subsection{Further research}

%%%%%%%%%%%%%%%%%%%%%%%%%%%%%%%%%%%%%%%%%%%%%%%%%%%%%%%%%%%%%%%%%%%

While our tools allow for the detection of loophole-free nonlocality by using only fractions of contexts (at the cost of a higher detection efficiency), when we apply these tools to the correlations in Fig.~\ref{fig:pnp_bell} and Table \ref{tab:graph_theoretic}, then the number of measurements required is either still too large for practical Bell tests or not too large but then they offer critical detection efficiencies that are comparable to the ones needed in existing detection-loophole-free experiments with smaller dimension and number of settings. In other words, the examples used in this work do not yet provide practical targets with sufficiently low critical detection efficiency and number of settings.

However, it is important to emphasize that the examples we have used to illustrate our tools are based on correlations obtained by sophisticated but arguably sub-optimal methods with the purpose of showing that the critical detection efficiency can be arbitrary low without the need of quantum systems of dimension impossible to achieve in the laboratory. Earlier it was shown that high-dimensional quantum systems can tolerate a detection efficiency that decreases with an increase in the local dimension $d$~\cite{M02}. However, an improvement over the qubits systems can only observed for $d>1600$. The correlations we study here require systems with significantly lower $d$. Presumably, there are many quantum correlations with sufficiently low critical detection and number of settings waiting to be discovered. Our hope is that the tools introduced here stimulate the search for such correlations.

%%%%%%%%%%%%%%%%%%%%%%%%%%%%%%%%%%%%%%%%%%%%%%%%%%%%%%%%%%%%%%%%%

\section*{Acknowledgments}

%%%%%%%%%%%%%%%%%%%%%%%%%%%%%%%%%%%%%%%%%%%%%%%%%%%%%%%%%%%%%%%%%%%

This work was supported by the Horizon 2020 \href{10.3030/101017733}{QuantERA II} project \href{https://quantera.eu/secret/}{SECRET} (Grant agreement No.\ 731473 and 101017733) by \href{10.13039/501100011033}{MCINN/AEI} (Project No.\ PCI2019-111885-2). A.C.\ was also supported by the Digital Horizon Europe project \href{10.3030/101070558}{FoQaCiA} (Grant agreement No.\ 101070558) and \href{10.13039/501100011033}{MCINN/AEI} (Project No.\ PID2020-113738GB-I00).

%%%%%%%%%%%%%%%%%%%%%%%%%%%%%%%%%%%%%%%%%%%%%%%%%%%%%%%%%%%%%%%%%%%

%\bibliography{references}


%%%%%%%%%%%%%%%%%%%%%%%%%%%%%%%%%%%%%%%%%%%%%%%%%%%%%%%%%%%%%%%%%%%

\end{document}